\documentclass[journal]{IEEEtran}

\usepackage{amsmath,graphicx}
\usepackage{tikz}
\usetikzlibrary{arrows,trees,positioning,shadows}
\usepackage{hyperref}
\usepackage{array}
\usepackage{bm}
\usepackage{amsfonts}
\usepackage{amsmath}
\usepackage{cite}
\usepackage{float}
\usepackage{tabularx}
\usepackage[caption = false]{subfig}
\usepackage[OT2,T1]{fontenc}
\DeclareSymbolFont{cyrletters}{OT2}{wncyr}{m}{n}
\DeclareMathSymbol{\sha}{\mathalpha}{cyrletters}{"58}
\ifCLASSINFOpdf
\else
\fi

\begin{document}
%
\title{Robust Downbeat Tracking Using an Ensemble of Convolutional Networks}
%
%
%

\author{Simon~Durand, 
        Juan~Pablo~Bello, 
        Bertrand~David, 
        and~Ga\"el~Richard
\thanks{S. Durand, B. David, and G. Richard are with the LTCI, CNRS, T\'el\'ecom ParisTech, Universit\'e Paris-Saclay, 75013, Paris, France.}
\thanks{J. P. Bello is with the Music and Audio Research Laboratory (MARL), New York University $-$ USA}}
\maketitle

\begin{abstract}
In this paper, we present a novel state of the art system for automatic downbeat tracking from music signals. The audio signal is first segmented in frames which are synchronized at the tatum level of the music. 
We then extract 
different kind of features based on harmony, melody, rhythm and bass content 
to feed 
convolutional neural networks that are adapted to take advantage of each feature characteristics. This ensemble of neural networks is combined to obtain one downbeat likelihood per tatum. The downbeat sequence is finally decoded with a flexible and efficient temporal model which takes advantage of the metrical continuity of a song. We then perform an evaluation of our system on a large base of 9 datasets, compare its performance to 4 other published algorithms and obtain a significant increase of 16.8 percent points compared to the second best system, 
for altogether a moderate cost in test and training. 
The influence of each step of the method is studied to show its strengths and shortcomings.
\end{abstract}


%
\IEEEpeerreviewmaketitle

\section{Introduction}
The time structure of a music piece is often conceived as a superposition of multiple hierarchical levels, or time-scales, interacting with one another~\cite{Lerdahl1983}. 
Automatically analysing and detecting those different temporal layers is thus of significant importance when we are trying to make sense of the data. 
When listening to a song, most people naturally synchronize to a specific level called the tactus or beat level~\cite{Large1999}. 
Depending on the duration, the loudness, the pitch of the events or even on the local prosody of the musical phrase, these beats are differently accented. It eventually leads to a grouping over a larger scale. This scale is, at least in the western music context, that of the bar which thus defines the metrical structure of the piece. 
The purpose of this work is to automatically estimate the locations of the first beat of each bar, the so-called downbeat, with the help of multiple features and deep neural networks especially designed for the task.  

Downbeats are often used by composers and conductors to help musicians read and navigate in a musical piece. 
Its automatic estimation is useful for various tasks such as automatic music transcription~\cite{Mauch20102}, genre recognition~\cite{Tsunoo2011}, chord recognition~\cite{Shenoy2005} and structural segmentation~\cite{Maddage2006}.
It is also useful for computational musicology~\cite{Hamanaka2014}, measuring rhythm pattern similarity~\cite{Paulus2002}, and synchronizing two musical excerpts~\cite{Hockman2008} or a musical piece with another media segment such as a virtual dancer, a drum machine, virtual books or a light show~\cite{Goto2001}. 

Downbeat tracking is a challenging task because it often relies on other subtasks such as beat tracking, tempo and time signature estimation and also 
because of the difficulty to state an unambiguous ground truth.  Current approaches are thus facing limitations in scope. 
For example, methods based on rhythmic features only are not applicable to a wide range of music styles 
since they implicitly relies on the presence of percussive instruments
~\cite{Krebs2013,Hockman2012,Srinivasamurthy2014}. Other systems 
choose to strongly limit the possible time-signature~\cite{Gartner2014,Klapuri2006,Goto2001,Davies2006,Papadopoulos2011DB,Peeters2011,Krebs2013,Hockman2012}, need downbeats to be at onset positions~\cite{Jehan2005}, or 
make use of restrictive prior knowledge
~\cite{Davies2006,Papadopoulos2011DB,Allan2004}. 
Approaches that estimate some necessary information beforehand are naturally prone to  error propagation. 
A number of current methods characterize downbeats with the help of one feature type only, while a more diverse description is proved to me more useful
\cite{Hockman2012,Goto2001,Peeters2011,Khadkevich2012}. Interestingly, there is an analogy with the multi-faceted aspect of 
human downbeat perception, which takes into account rhythm, but also harmony, musical structure and melodic lines \cite{Lerdahl1983}. 
At last,  downbeat detection functions are often 
computed 
from low-level features, without 
taking into account higher-level representations. 
Some are using  such representations, as in \cite{Papadopoulos2011DB} but in those cases, the algorithms rely on a chord classification which could itself account for some errors. 




We therefore propose an approach that:
\begin{itemize}
\item Minimizes assumptions in feature, classifier and temporal model design, and therefore is more generalizable.
\item Does not require any prior information. 
\item Combines features of different kind to cover different aspects of the musical content.
\item Uses deep learning to obtain higher-level representations that fully characterize the complexity of the problem but are hard to design by hand.
\end{itemize}
The model is evaluated on a large number of songs from various music styles and 
shows a significant improvement over the state of the art. 

Subsequently, the paper is organized as follows. Section~\ref{sec:sota} 
states the problem of downbeat tracking and presents some of the related work. Section~\ref{sec:model} provides an overview of the method 
and  
its main steps 
Section~\ref{sec:dnnLearn} describes the 
neural networks and their tuning together with the learning strategies. Finally, section~\ref{sec:results} provides 
the experimental results and a comparison with four other published downbeat tracking systems. 

\section{Related Work}
\label{sec:sota}

\begin{figure}

\begin{center}

\includegraphics[width=9.5cm]{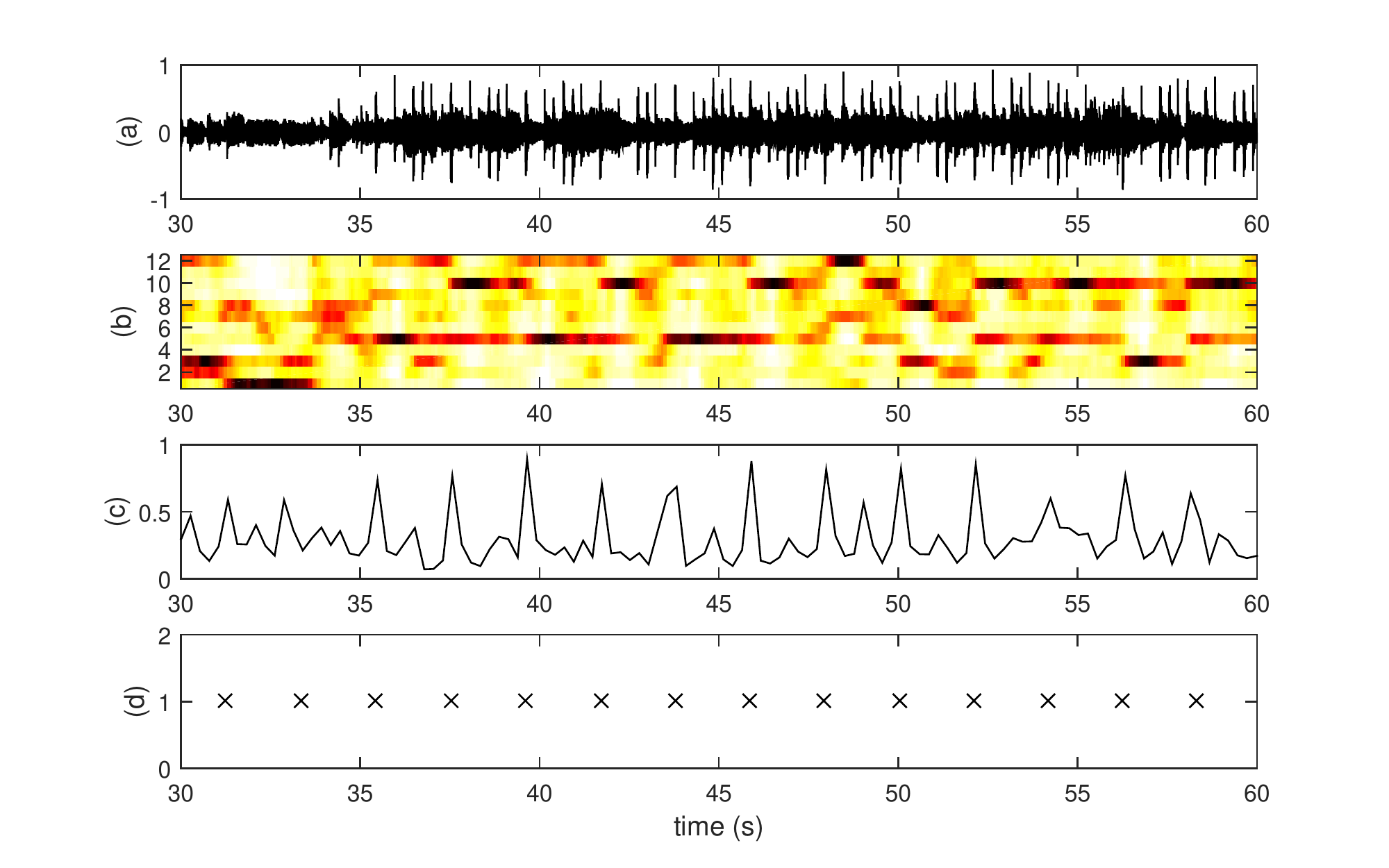}
\end{center}
\caption{\footnotesize Illustration of the three-step process of downbeat tracking for thirty seconds of audio. (a) Audio signal $\textbf{x}$. (b) A feature $\textbf{F}_i$, in this case chroma vectors. (c) Downbeat detection function \textbf{d}, in this case downbeat likelihood. (d) Discrete downbeat sequence \textbf{s}.}
\label{fig:Overview_DB}
\end{figure}

Algorithms for downbeat tracking are designed to estimate a discrete sequence of time instants corresponding to the bar positions in a musical piece.  In most cases, they can be divided into three mains steps, as illustrated in figure~\ref{fig:Overview_DB}:
\begin{enumerate}
\item From raw data compute feature vectors or matrices.
\item From features  derive a function representing the downbeat likelihood (the so-called downbeat detection function).
\item Obtain the downbeat sequence with the help of a temporal model.
\end{enumerate}

A wide range of techniques and ideas have been tested for each step, which are summarized in the following.
\subsection{Feature extraction}
\label{ssec:featExtrac}
Past approaches use domain-specific knowledge of music for feature design, often from a single attribute only. 
For example, chromas~\cite{Papadopoulos2011DB,Peeters2011,Hockman2012,Khadkevich2012} and spectral component histograms~\cite{Goto2001} have been utilized as harmonic features, having in mind that the harmonic content is more likely to change around the downbeat positions.  

Other features, more generally considered as rhythm markers, are based on the onset detection function (ODF)~\cite{Goto2001,Gainza2007,Whiteley2006,Hockman2012,Khadkevich2012}. 
It is worth noting here that, even if isolated onsets are not always present at downbeat positions, onset patterns will often be synchronous with the bar changes. 
ODF are often extracted across frequency bands~\cite{Krebs2013,Holzapfel2014db} in order, for instance, to separate the events corresponding to kick drums or bass  from those of the snare.  

A third category concerns timbre inspired features~\cite{Srinivasamurthy2014,Hockman2012,Jehan2005,Durand2014}. 
Alterations of the timbre content occur more likely at the start of a new section and near a downbeat position. 
This feature extraction step can be done in conjunction with an onset~\cite{Jehan2005}, tatum~\cite{Gartner2014} or beat~\cite{Davies2006} segmentation. 

\subsection{Downbeat detection function}
The goal of the second step is to map the features into a downbeat detection function. This can be done with heuristics. 
When harmony features are used, most systems focus on measuring a change in the feature properties~\cite{Peeters2011,Khadkevich2012,Davies2006,Goto1999}. 
This can be done with the cosine distance, the Kullback-Leibler divergence or a difference of the principal harmonic components.
If rhythm features are considered, comparison with pre-stored rhythm patterns~\cite{Goto2001,Klapuri2006,Whiteley2006}, beat enhanced onset detection function~\cite{Hockman2012} 
and relative spectral balance between high and low energy content at different beat positions~\cite{Peeters2011} have also been proposed.

Machine learning systems can also be considered. Generic approaches such as
Support Vector Machines (SVM) with an auditory spectrogram~\cite{Jehan2005,Hockman2012} have been used at the risk of not being tailored for this particular problem. 
Other systems focus on recognizing rhythm patterns in the data to find the downbeats. It can be done with a Conditional Deep Belief Network~\cite{Battenberg2012} (CDBN) or a Gaussian Mixture Model (GMM) coupled to a k-means clustering
~\cite{Krebs2013,Holzapfel2014db}. These rhythm related methods are more adapted to the problem but they often make strong style-specific assumptions and require music with a very distinctive rhythm style to work well~\cite{Krebs2013,Holzapfel2014db,Rocamura2016}.

\subsection{Downbeat sequence extraction}
The goal of the last step is to discretize the downbeat detection function into a downbeat sequence. Considering that the temporal distance between two downbeats varies slowly, it can be useful to estimate the bar periodicity beforehand. To do so, it is possible to segment the audio into frames of different lengths and find the length that makes those segments the most similar~\cite{Gainza2007}. We can also take advantage of the repetitive aspect of the onset detection function with a comb filter or a short time Fourier transform to find different periodicity levels: \textit{tatum}, \textit{tactus}, and \textit{measure}. Assuming an integer ratio between these levels can help estimating them jointly~\cite{Klapuri2006} or successively~\cite{Gartner2014,Srinivasamurthy2014}.

To take into account the slow bar length variation over time, induction methods are often used. The estimated downbeat sequence will be the one that maximizes the downbeat detection function and minimize the bar length variation. It can be done in a greedy way, one downbeat after another, starting at the first downbeat~\cite{Goto2001,Gainza2007}
 or at the start of the first characteristic rhythm pattern~\cite{Gartner2014}. To avoid being stuck in a local minimum, most algorithms search a more global downbeat sequence path. It can be done with dynamic programming~\cite{Hockman2012} or Hidden Markov Models (HMM)~\cite{Klapuri2006,Papadopoulos2011DB,Peeters2011} that can sometimes be coupled with a learned language model~\cite{Khadkevich2012}. A particularly interesting temporal model is the dynamic bar pointer model~\cite{Whiteley2006}. It consists of a dynamic bayesian network jointly modeling the downbeat positions, the tempo and the rhythm pattern. 
This method was refined to improve the observation probabilities~\cite{Krebs2013} 
and reduce of the computational complexity of the inference~\cite{Krebs2015}. Most of the aforementioned systems don't handle varied time signatures during the downbeat sequence extraction.

\begin{figure*}

  \centering
\includegraphics[width=17.8cm]{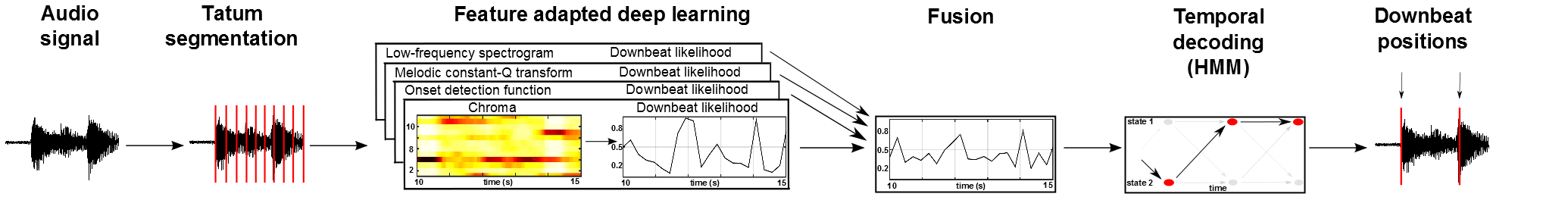}

\caption{\footnotesize Model overview. }
\label{fig:ModelOverview}
\end{figure*}
\section{Proposed model}
\label{sec:model}
We will describe in this section the five parts of the proposed model, after a general overview.

\subsection{Model overview}

The model overview is illustrated in figure~\ref{fig:ModelOverview}. The signal's time-line is quantized to a set of bar subdivisions so-called tatums. The purpose of the system is to classify those tatums as being or not at a downbeat position. Features related to harmony, rhythm, melody and bass content are computed from the signal and mapped to a finer temporal scale. Inputs centered around each candidate tatum are extracted. Each input is then fed to an independent deep neural network (DNN). The DNNs either classify all the input tatums or just the tatum at the middle of the input to be at a downbeat position or not and output a downbeat likelihood function. Networks outputs are fused to obtain a single downbeat likelihood per tatum. Finally, a HMM is used to estimate the most likely downbeat sequence.

\subsection{Tatum synchronous segmentation}

We adapt the local pulse information extractor proposed in~\cite{tempogram2011} to achieve a useful tatum segmentation. 
The first step is to compute the tempogram of the musical audio through a short-term Fourier transform (STFT) and to only keep the tempo above 60 repetitions per minute to avoid slow metrical levels. We then track the best periodicity path by dynamic programming with the same kind of local constraints as in~\cite{Alonso2006}\footnote{They are more permissive in our case, [0.5, 0.7, 1 0.7, 0.5] instead of [0.95 0.98 1 0.98 0.95], to be more flexible to moderate tempo change. See~\cite{Alonso2006} for more detail on the dynamic programming computation.}.  The following system can find a fast subdivision of the downbeats at a rate that is locally regular. We finally use the decoded path to recover instantaneous phase and amplitude values, construct the predominant local pulse (PLP) function as in~\cite{tempogram2011}, and detect tatums using peak-picking on the PLP. The resulting segmentation period is typically twice as fast as the beats period, while it can be up to four times faster. 

It is interesting to use tatums\footnote{It is to note that we are not formally looking for tatums, but more precisely for a regular and fast downbeat subdivision. The result is close to tatums and is called this way for convenience.} as those events for several reasons. First, it encodes a musically meaningful dimension reduction based on tempo invariance as opposed to short fixed frames. This also lowers the cost of designing, training and testing deep neural networks and probabilistic temporal methods. Finally, it is possible to design a tatum segmentation method with high recall rate, meaning that almost all possible downbeats are candidate for detection.

\subsection{Feature extraction}
In this work, the aim of feature extraction is to represent the signal as a function of four musical attributes contributing to the grouping of beats into a bar, namely harmony, rhythmic pattern, bass content, and melody. This multi-faceted approach is consistent with well-known theories of music organization \cite{Lerdahl1983}, where the chosen attributes contribute to the perception of downbeats. 
In section~\ref{ssec:featExtrac}, we discussed why harmony and rhythm features are useful for downbeat tracking. 
 The bass or low-frequency content contains mostly bass instruments or kick drum, both of which tend to emphasize the downbeat. 
For melody, some notes tend to be more accented than others and both pitch contour and note duration play an important role in our interpretation of meter \cite{Ellis2009,Hannon2004,Pfordresher2003}. 
These attributes will be represented by chroma, low-frequency spectrogram (LFS), onset detection function (ODF) and melodic constant-Q transform (MCQT) features respectively. Each representation, illustrated in figure~\ref{fig:net_ic16}, is computed from a STFT using a Hann window of varying size applied to a resampled input signal as given in Table~\ref{tab:feat_params}. More details on their implementation are provided below.
\begin{table}
	\centering
	\caption{STFT analysis parameters for each representation. $s_r$ is the sampling rate used in Hz. Sizes are given in ms.}
	\begin{tabular}{cccc}
		  & Window size & Hop size & $s_r$  \\ 
		\hline\hline Chroma & 743 & 92.2 & 5512.5 \\ 
        LFS & 64 & 8 & 500 \\
		ODF & 23.2 & 11.6  & 44100 \\ 		 
		MCQT & 185.8 & 11.6  &  11025 \\ 
		
	\end{tabular} 
	\label{tab:feat_params}	
\end{table}
\subsubsection{Chroma}
The chroma computation is done as in \cite{Bello2005chro}. 
We apply a constant-Q filter-bank with 108 bins (36 bins per octave) to the STFT coefficients and convert the constant-Q spectrum to harmonic pitch class profiles. Octave information is removed by accumulating the energy of equal pitch classes. The chromagram is tuned and then smoothed by a median filter of length 8. It is finally mapped to a 12 bins representation by averaging.


\subsubsection{Low-frequency spectrogram}
We only keep spectral components of the STFT representation below 150 Hz (the first 10 bins). To limit the variation of this feature, the signal is clipped so that all values above the $9^{th}$ decile are equal.

\subsubsection{Onset detection function}

We compute a three bands spectral flux ODF. To do so, we use $\mu$-law compression, with $\mu=10^6$ to the STFT coefficients. We then sum the discrete temporal difference of the compressed signal on three bands for each temporal interval, subtract the local mean and half wave rectify the resulting signal. The frequency intervals of the low, medium and high frequency bands are [0 150], [150 500] and [500 11025] Hz respectively as we believe low frequency bands carry a lot of weight in our problem. 
To limit the variation of this feature, as for the LFS, the signal is clipped so that all values above the $9^{th}$ decile are equal. 

\subsubsection{Melodic constant-Q transform}


We apply a constant-Q transform (CQT) with 96 bins per octave, starting from 196 Hz to the Nyquist frequency to the STFT, and average the energy of each CQT bin $q[k]$ with the following octaves: \vspace{-6pt}
\begin{equation}
q_a[k] = \frac{\sum_{j=0}^{J_k}q[k+96j]}{J_k+1}
\end{equation}
with $J_k$ such that $q[k+96J_k]$ is below the Nyquist frequency. We then only keep 304 bins from 392 Hz to 3520 Hz that correspond to three octaves and two semitones. The lower frequency is significantly higher than the chroma feature lower frequency and allows more diversity with this feature and the range is wide enough to capture most of the melodic lines while keeping a relatively low computational cost. We use a logarithmic representation of $q_a$ to represent the variation of the energy more clearly: 

\begin{equation}
lq_a = \log(\left|\hat{q_a}\right| + 1)
\end{equation} 
where $\hat{q_a}$ is the restriction of $q_a$ between 392 Hz and 3520 Hz. Additionally, we zero all values below the third quartile $Q_3$ of a given temporal frame to get our melodic CQT: 
\begin{equation}
m_{CQT} = \max(lq_a - Q_3(lq_a),0)
\end{equation}
Keeping only the highest values allows us to remove most of the noise and onsets so we can see some contrast and obtain features that are significantly different from the previous rhythm feature. 

\subsubsection{Temporal quantization}
The aforementioned features are then mapped to a grid with subdivisions lasting one fifth of a tatum using interpolation. We therefore have tempo independent features with a temporal precision higher than the tatum level. Their temporal or horizontal dimension is therefore 5 per tatum and their vertical dimension is respectively 12, 10, 3 and 304. We segment these features so they start at the beginning of each tatum and have a fixed length of 9 or 17 tatums. They are finally scaled between 0 and 1 and act as input to the deep neural networks as described in the following section.

\subsection{Feature learning}
\label{cnns}
Downbeats are high-level constructs depending on complex patterns in the feature sequence. We propose that the probability of a downbeat can be estimated using feed-forward DNN $F(X_0 | \Theta,P)$, where $X_0$ is our input tensor of temporal, spectral and feature map dimension $[N_0,M_0,L_0]$, and $\Theta$ and $P$ are the learned and designed parameters of the network. In our implementation, $F$ is a cascade of $K=4$ non-linear functions $f_k(X_k | \theta_k, p_k)$, with $ \theta_k = [W_k,b_k]$, where $W_k$ is a tensor of weights, $b_k$ is a vector of biases, and the tensor $X_k$ is the output of layer $k-1$ for $k > 0$, and the input feature for $k = 0$. $f_k$ will be a cascade of a convolution $c$ and one or several non linear functions $h_k$:
\begin{equation}
f_k(X_k|\theta_k,p_k) = h_k(c(X_k,\theta_k,p_{1_k}),p_{2_k});\; \forall k \in [0..K-1]
\end{equation}
In our case, $p_{1_k}=[t_{1_k},v_{1_k},L_k,n_{1_k}]$ with $t_{1_k}$ and $v_{1_k}$ the temporal and spectral dimensions of $W_k$, $L_k$ the depth of $X_k$, and $n_{1_k}$ the number of filters. $c$ is defined as:
\begin{equation}
c = b_k[l'] + \sum_{t=1}^{t_{1k}}\sum_{v=1}^{v_{1k}}\sum_{l=1}^{L_{k}}W_k[t,v,l,l']X_k[t'+t-1,v'+v-1,l]\\
\label{eq:convo}
\end{equation}
where $t' \in [1..N_{k+1}]$, $v' \in [1..M_{k+1}]$ and $l' \in [1..n_{1_k}]$. $h_k$ is in our case a set of one or several cascaded non linear functions among rectified linear units $r(x)=\max{(0,x)}$ \cite{Zeiler2013}, sigmoids $\sigma(x)=\frac{1}{1-e^{-x}}$, max pooling $m$, softmax normalization $s(x)=\frac{e^{x}}{\sum_{j=1}^{J}e^{x[j]}}$ and dropout regularization $d$ \cite{Hinton2012}. $p_{2_k}=[t_{2_k}, v_{2_k}]$ is the designed set of parameters of $h_k$ corresponding in our case to the temporal and vertical dimension reduction of the max pooling. $X_K$ will be the final output and will act as a downbeat likelihood, and sometimes its complementary.

In this work, we consider each feature type independently and we train one DNN per feature. This is a convenient way to work with features of different dimension and assess the effect of each of them. Moreover, we want to adapt feature learning to the feature type. Harmonic features will mostly exhibit change while rhythm features will often show characteristic patterns as downbeat cues. Low-frequency and melodic features will exhibit a bit of both and melodic features need an adapted dimensionality reduction process for example. Besides these differences, there is no reason that the optimal DNN hyper-parameters have to be the same in each case. The specific adaptations are described in section~\ref{sec:dnnLearn}. We use the MatConvNet toolbox to design and train the networks \cite{Vedaldi2014}. 

\subsection{Feature combination}
For each tatum we have four intermediary downbeat likelihoods. We need to fuse this information into a single robust downbeat likelihood to feed our temporal model. We will use for that purpose the results average. The average, or sum rule is indeed in general quite resilient to estimation errors \cite{Kittler1998}. However, it is not robust to redundant information and the network will need to produce complementary information.

\subsection{Temporal modeling}
\label{ssec:Viterbi}
We use a first order left-to-right HMM to map the continuous downbeat likelihood function $\mathbf{d}$ into the discrete sequence of downbeats $\mathbf{y}$. The inference is done with the Viterbi algorithm and the temporal interval of our model is the tatum. Our model contains:
\begin{itemize}
\item The state space $H=\{1...N_h\}$ with $N_h$ the number of hidden states $i$.

Since the downbeat likelihood depends on the bar length and the position inside a bar, we will define a state for each possible tatum in a given bar. 
For instance, considering two possible bars of two and three tatums, there would be five different states in the model. One state would represent the first tatum of the two-tatum bar, another would represent the second tatum of the two-tatum bar and so on. In practice, we allow time signatures of \{3,4,5,6,7,8,9,10,12,16\} tatums per bar, for a total of $N_h=80$ states. Furthermore, modeling bars of different length independently allows for consistency in the decoding stage. We can enforce that bars with the same number of tatum are found for example.

\item The most likely state sequence $\mathbf{y'}= [ y'(1);...;y'(T)]$ with $y'(k)\in H,\;\forall k \in \{1...T\}$ and $T$ the length of the sequence.

\item The initial probability $\pi_i=P(y'(1)=i)$ of being in a state $i$ initially.

We use an equal distribution of the initial probabilities: $\pi_i=\frac{1}{N_h},$ $\forall i \in H$, for robustness.

\item The observation sequence $\mathbf{o}= [ o(1);...;o(T)]$ with $o(k)\in[0,1],\;\forall k \in \{1...T\}$.

It is equal to the downbeat likelihood: $\mathbf{o}=\mathbf{d}$. 

\item The emission probabilities $\mathbf{e}_i= [ e_i(1);...;e_i(T)]$ with $e_i(k)=P(o(k)|y'(k)=i)$ the probability of the observation $o(k)$ given a state $i$, $\forall k \in \{1...T\}$.

For the emission probabilities we will distinguish two cases, either the state corresponds to a tatum at the beginning of a bar: $i\in H_1\subset H$, either it corresponds to another position inside a bar: $i\in\overline{H}_1\subset H$. In the first case the emission probability is equal to the downbeat likelihood and in the second case to its complementary probability:
\begin{equation}
\mathbf{e}_i = \begin{cases}
\mathbf{d}  &\text{if $i \in H_1$}\\
1-\mathbf{d} &\text{if $i \in \overline{H}_1$}
\end{cases}
\end{equation}
\item The transition probability $\mathbf{a}_{i,j}= [ a_{i,j}(2);...;a_{i,j}(T)]$ with $a_{i,j}(k)=P(y'(k)=j|y'(k-1)=i)$ the probability of transitioning from state $i$ to state $j$, $\forall k \in \{2...T\}$.

The transition probabilities will only depend on the preceding and current state: $a_{i,j}(k)=a_{i,j},\; \forall k \in \{2...T\}$. They are therefore defined by a transition matrix $\mathbf{A}=\{a_{i,j},\; (i,j)\in\{1...N_s\}^2\}$. The ${N_s}^2=6400$ parameters of our transition matrix could be trained entirely automatically but this is left for future work. We train the majority of transition matrix parameters by simply counting the number of occurrences for a specific transition and giving a minimum value if an occurrence didn't sufficiently happen. If a transition from $i$ to $j$ occurs $n$ times out of a total of $N$ transitions from $i$ to any state, then $a_{i,j} = \max\left(\frac{n}{N},0.02\right)$. Putting non null values to $a_{i,j}$ allows for a better adaptability to the downbeat likelihood from the model. Finally, some key transitions are manually fine-tuned by maximizing the F-measure of the training set. It appeared that over-fitting was not an issue for this problem, probably because of a relatively wide range of close to optimal values. 
To give an idea, the transition matrix coefficients can be summarized in three categories. There are high probabilities to advance circularly in a given bar, medium probabilities to go from the end of one bar to the beginning of another, and low probabilities to go elsewhere.
 \end{itemize}
 
In the end, decoded states corresponding to a tatum at the beginning of a bar will give the downbeat sequence: $\mathbf{y} = \{\mathbf{y'},\; y'(k)\in H_1 \}$.

\section{Feature adapted deep neural networks}
\label{sec:dnnLearn}
To take advantage of the specificities of the different extracted features, we first exploit their local structure by using convolutional neural networks (CNN). In CNNs, convolutional layers \cite{Lecun1998,Lecun2010} share weights across the inputs within a spatio-temporal region so each input unit is not considered independent from its neighbors. 
We then adapt the architecture and the learning strategies to each input as described below. 

\subsection{Melodic neural network (MCNN)}
\begin{figure}

  \centering
\includegraphics[width=8.3cm]{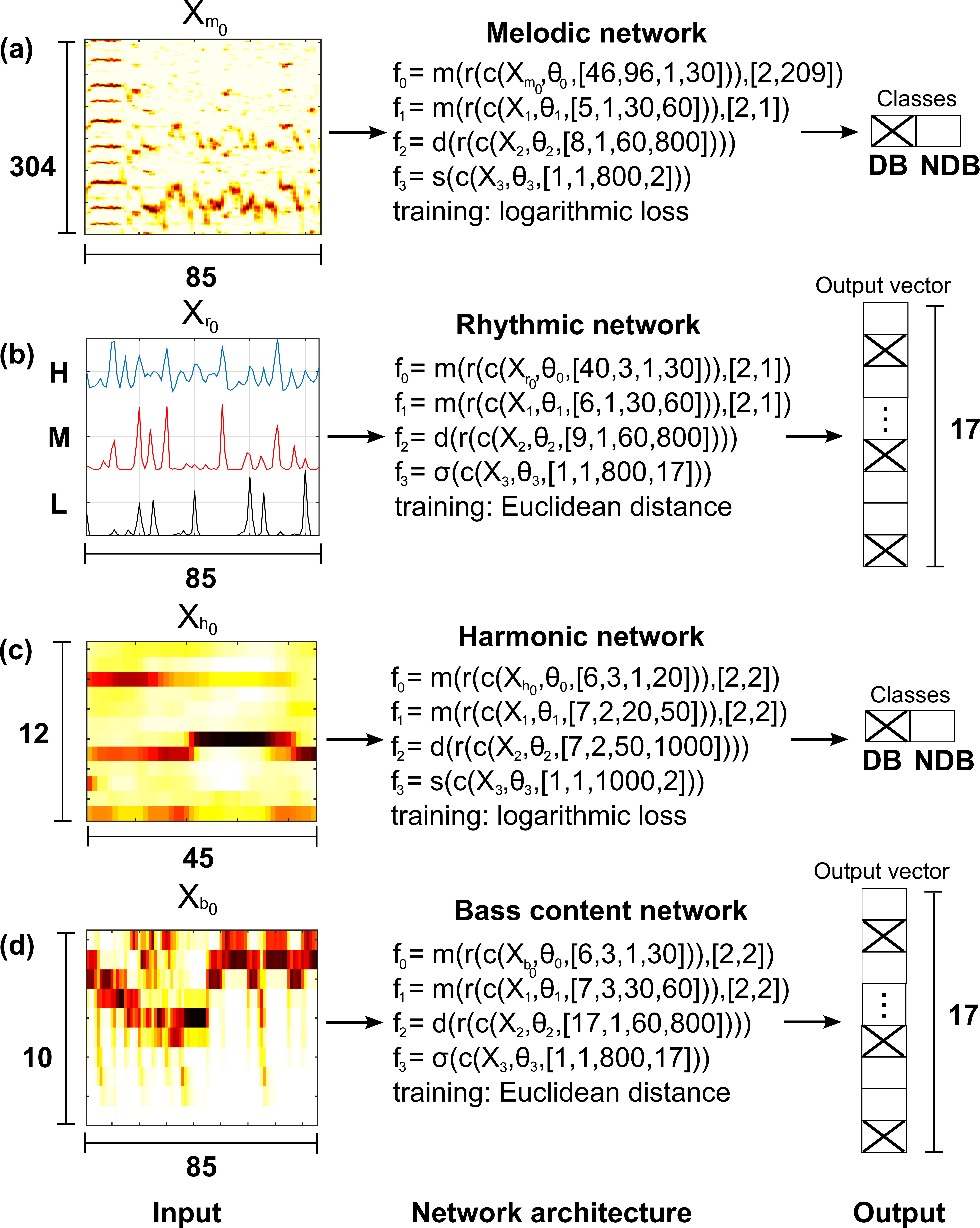}
\caption{\footnotesize Convolutional networks architecture, inputs and outputs. The notation is the same as in subsection~\ref{cnns}. DB and NDB stand for downbeat and no downbeat respectively.}
\label{fig:net_ic16}
\end{figure}

The goal of the melodic network is to learn melodic contours as they play a role in our meter perception regardless of their absolute pitch~\cite{Thomassen1982}. Considering that those patterns can be relatively long, we will use 17 tatum-long inputs. It roughly corresponds to two bars of audio in 4/4. We then have input features $X_{m_0}$\footnote{The index indicating the type of network, here $m$ for melodic network, won't be explicitely mentioned in the following for simplicity.} of spectral dimension $M_0=304$ and of temporal dimension of 17 times 5 temporal quantization: $N_0=85$. Our network architecture is presented in figure~\ref{fig:net_ic16}(a). For example, the first layer:
\begin{equation}
f_0=m(r(c(X_0,\theta_0,[46,96,1,30])),[2,209])
\end{equation}  
means that we use filters of parameters $\theta_0$ and size $[46,96,1,30]$ on input $X_0$ for convolution $c$, and then rectified linear units $r$ and max pooling $m$ with a reduction factor of $[2,209]$ as non linearity.
The first layer filters are relatively large, $t_{1_0}=46$ and $v_{1_0}=96$, so we are able to characterize pitch sequences. The reduction factor in frequency of the following max pooling, $v_{2_0}=209$, is equal to the input size after the convolution. This way, we only keep the maximal convolution activation in the whole frequency range and lead the network to focus on patterns regardless of their absolute pitch center. The fourth layer, $f_3=s(c(X_3,\theta_3,[1,1,800,2]))$, can be seen as a fully connected layer: $t_{3_0}=1$ and $v_{3_0}=1$, that will map the preceding hidden units into 2 final outputs. Those outputs represent the likelihood of the center of the input to be at a downbeat position and its complementary. We train the network with the logarithmic loss between the outputs and the ground truth.

%

\subsection{Rhythmic neural network (RCNN)}
The rhythmic network aims at learning rhythmic patterns of specific length, instead of sudden changes in the ODF that are not necessarily indicative of a downbeat position. Since rhythm patterns can be long we also use 17 tatum long inputs. 
Contrary to melodic patterns, the length of a rhythmic pattern and the length of a bar are strongly correlated and the pattern boundaries are likely to be located at a downbeat position. To characterize this pattern length, the network should give different outputs if patterns of different length are observed. We choose for that multi-label learning~\cite{Tsoumakas2006}. In this case, if there is a downbeat position at the first and ninth tatum of our 17 tatum-long input, the output of our network should be $X_4=[1\: 0\: 0\: 0\: 0\: 0\: 0\: 0\: 1\: 0\: 0\: 0\: 0\: 0\: 0\: 0\: 0]$. Since there might be multiple downbeats per input, it is not appropriate to normalize the network output $X_4$. Instead, we want each coefficient of $X_4$ to be close to 0 if there isn't a downbeat and 1 if there is a downbeat at the corresponding tatum position. Therefore, we first use a sigmoid activation $\sigma$ unit in our last layer to map the results into probabilities: $f_3=\sigma(c(X_3,\theta_3,[1,1,800,17]))$. We then train the network with an Euclidean distance between the output and the ground truth $\mathbf{g}$ with a similar shape as $X_4$: $g(k)=1$ if there a downbeat at the $k^{th}$ tatum of the input and $g(k)=0$ otherwise. Each tatum is then considered independent. Our network architecture is presented in figure~\ref{fig:net_ic16}(b). To detect bar-long rhythm patterns, our first convolutional layer uses relatively large temporal filters of about the length of a bar,  $t_{1_0}=40$ and $v_{1_0}=3$. 
Besides, since we are using the Euclidean distance to ground truth vectors to train the network, we are not explicitly using classes such as 'downbeat' and 'no downbeat'. The output is then of dimension $17$ and represents the downbeat likelihood of each tatum position in ${X_r}_0$. Since we have 17 tatum-long inputs but a hop size of 1 tatum, overlap will occur. We will reduce the dimension of our downbeat likelihood to 1 by averaging the results corresponding to the same tatum.

\begin{figure}

  \centering
\includegraphics[width=9.1cm]{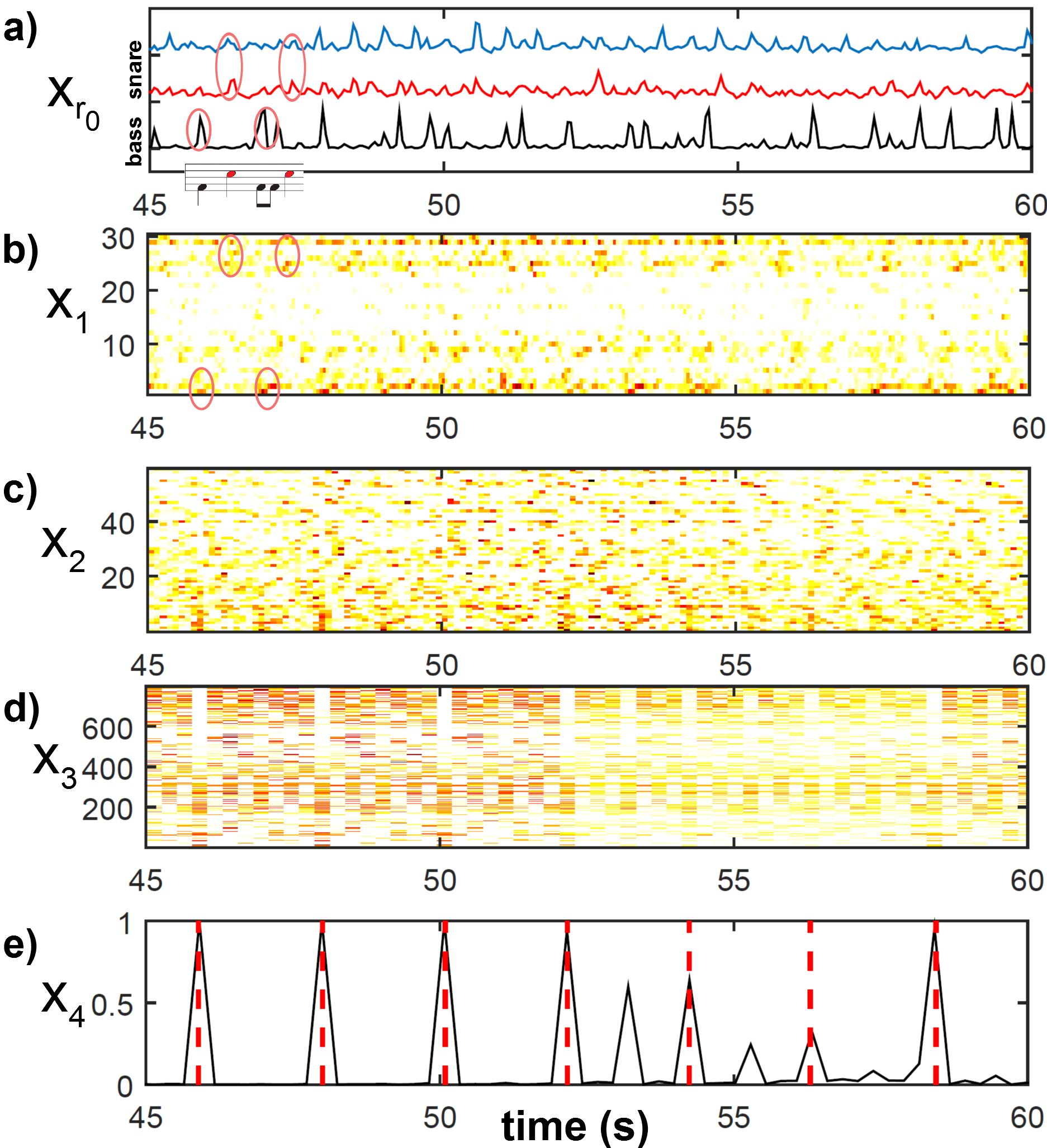}
\caption{\footnotesize RCNN input and intermediary and final outputs. a): Onset detection function input. b-c): Output of all the filters on top of each other, from the first and second layer respectively before the max pooling. d): Output of all filters from the third layer. e): Output of the network, that acts as a downbeat likelihood. The red dotted lines are the annotated downbeat positions. Since inputs overlap in time, the figures are averaged to get one bin per filter per time frame when necessary.} 
\label{fig:rcnn}
\end{figure}

Figure~\ref{fig:rcnn} represents the input transformation through the network until the downbeat likelihood. It can be seen that in the first layer some units tend to be activated around rhythmic patterns or events, highlighted here in part by the orange circles in figure~\ref{fig:rcnn}b). As we go deeper into the network, some units tend to be activated more clearly around downbeat positions and some other units around no-downbeat positions. We finally obtain a rather clean downbeat likelihood function, in figure~\ref{fig:rcnn}e), that is high around the red dotted lines representing the annotated downbeats. The network is a bit more indecisive around the fifty fifth second, as there is a drum fills leading to a chorus.



\smallskip
\subsection{Harmonic neural network (HCNN)}
The goal of this network is to learn how to detect harmonic changes in the input (see figure~\ref{fig:net_ic16}(c)). Filter and input temporal size are then rather small. The input temporal dimension is only of 9 tatums, $N_0=45$, and the first convolutional layer contains filters of moderate size, $t_{1_0}=6$ and $v_{1_0}=3$ for example. 
Here, we do not need to characterize the length of a pattern and we then choose the same kind of non-linear functions as in the melodic network for the four network layers and logarithmic loss to train the network. The size of the filters differs to be adapted to the input size though. 
Additionally, it is desirable for this network to be robust to song transposition, as it changes the input but not our downbeat perception. To that aim, max pooling on the whole frequency range as in the melodic network will not work because chroma vectors are circular. Instead we choose to augment the training data with the 12 circular shifting combinations of the chroma vectors. 

\begin{figure}

  \centering
\includegraphics[width=9.1cm]{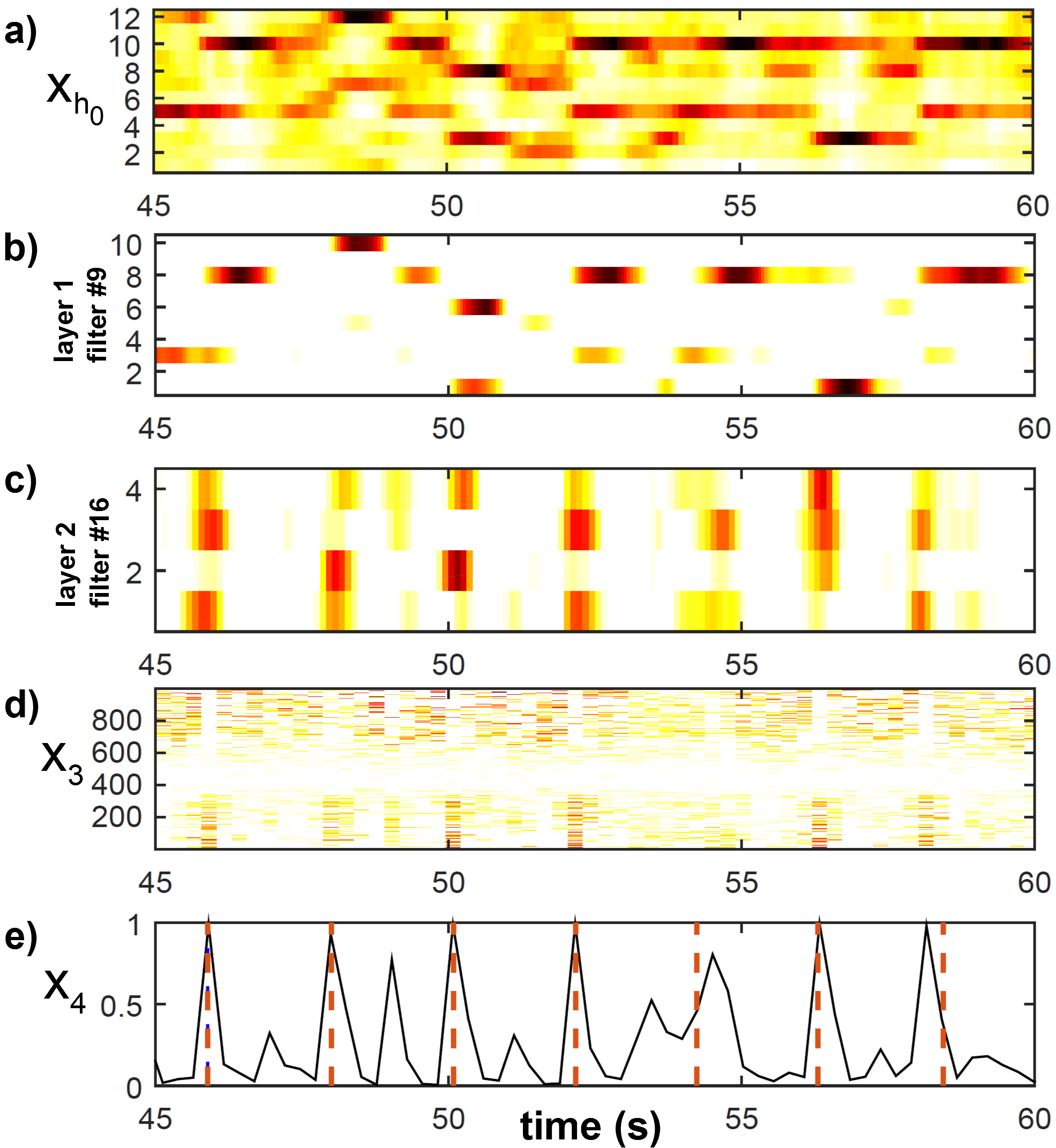}
\caption{\footnotesize HCNN input and intermediary and final outputs. a): Chroma input. b): Output of the filter number 9 of the first layer before max pooling. c): Output of the filter number 16 of the second layer before max pooling. d): Output of all the filters of the third layer. e): Output of the first class of the network, that acts as a downbeat likelihood. The red dotted lines are the annotated downbeat positions. Since inputs overlap in time, the figures are averaged to get one bin per filter per time frame when necessary.} 
\label{fig:hcnn}
\end{figure}

The network input, layers output and final output can be seen in figure~\ref{fig:hcnn}. The first layer of the network transforms the input to remove some of the noise as in figure~\ref{fig:hcnn}b) or the highlight some of its properties such as its onsets or offsets for example. The second layer takes advantage of this new representation to reduce its dimension or compute some sort of harmonic change. The third layer outputs 1000 units of dimension 1 and we can see that some of them act as downbeat detectors, lower figure~\ref{fig:hcnn}d), and some of them as no-downbeat detectors, upper figure~\ref{fig:hcnn}d). Finally, the obtained downbeat likelihood, although a little noisy, is rather close to the ground truth downbeats.

\smallskip
\subsection{Bass content neural network (BCNN)}
The bass content feature contains melodic and percussive bass instruments as can be seen in the figure~\ref{fig:net_ic16}(d) by the horizontal and vertical lines respectively. Our network architecture is also presented in figure~\ref{fig:net_ic16}(d). Detecting patterns is useful here but instantaneous events are more directly related to downbeats than for the melodic feature for example as bass notes or bass drums tend to be played at a downbeat position. We therefore use filters of moderate temporal size for our first layer to emphasize these events, $t_{1_0}=6$. As bass events may be repeated each bar, we want to be able to characterize the pattern length like with the rhythmic network. We therefore use the same last layer architecture: $f_3=\sigma(c(X_3,\theta_3,[1,1,800,17]))$, and multi-label procedure with the Euclidean distance to a ground truth vector $\mathbf{g}$ of zeros around no downbeat and ones around downbeats as a cost function to minimize. The dimension of our downbeat likelihood function will also be reduced to one by averaging.

\section{Evaluation and results}
\label{sec:results}
The proposed system is compared to 4 other published downbeat tracking algorithms on a total of 9 datasets presented below. We also assess each step of our method and present some of its limitations and strengths.

\subsection{Methodology}
\subsubsection{Evaluation metrics and procedure}
We use the F-measure, expressed in percent, to evaluate the performance of our system. This widely used metric~\cite{Papadopoulos2011DB, Khadkevich2012,Peeters2011} is the harmonic mean of precision and recall rates. We use a tolerance window of $\pm$70 ms and do not take into account the first 5 seconds and the last 3 seconds of audio as annotations are sometimes missing or not always reliable. To assess statistical significance, we perform a Friedman's test and a Tukey's honestly significant criterion (HSD) test with a 95\% confidence interval. The best performing method and the ones with non statistically significant decreases will be highlighted in bold in the result Tables. Finally, to be fair to methods were not originally trained in all datasets, we use a leave-one-dataset-out approach as recommended in~\cite{Livshin2003}, whereby in each iteration we use all datasets but one for training and validation, and the holdout dataset for testing. 

\subsubsection{Datasets}
\label{ssec:datasets}
We evaluate our system on nine different real audio recording datasets, presented in Table~\ref{dataset_table}. The letter "e" in the "\# Tracks" column means that the tracks are excerpts of about 1 minute, compared to full songs elsewhere. Those datasets, while being mainly focused on western music, cover a wide range of styles, including pop, classical, jazz, choral, hiphop, reggae, disco, blues, electro, latin, rock, vocal and world music. They include more than 1500 audio tracks ranging from 30 seconds excerpts to 10-minute long pieces for a total of about 43 hours of music and 78171 annotated downbeats. Long excerpts will be more sensitive to an adapted temporal model while an instantaneous downbeat estimation is more important for short excerpts. Some datasets focus on a certain music style or an artist while others include a wider musical spectrum and are labelled as "Various" in the Table~\ref{dataset_table}. The subset of the Klapuri dataset gathers 40 randomly selected tracks among four music styles - Blues, Classical, Jazz and Electro/Dance - given two constraints : the time signature is fixed for each excerpt and there are 10 tracks per genre. 
Besides variety in content, using several datasets creates variety in annotation. Downbeats being an abstract and perceptive concept, different annotations will occur depending on the annotator, or if an automatic software was used beforehand for example.

{\footnotesize
\begin{table}
\begin{center}
\caption{\footnotesize Datasets overview.}
{\begin{tabular}{ lllll}
Name & {\# Tracks} & {\# DB} & {Length} & {Genre} \\
\hline \hline
RWC Class \cite{Goto2002rwc} & $60$& 10148 & 5h 24m & Classical \\
Klapuri subset \cite{Klapuri2006}& $40 - e$& 1197 & 0h 38m & Various \\
Hainsworth \cite{Hainsworth2004}&$222 - e$& 6180 & 3h 19m & Various\\
RWC Jazz \cite{Goto2002rwc}&$50$& 5498 & 3h 44m & Jazz\\
RWC Genre \cite{Goto2003rwc}&$92$& 11053 & 6h 22m & Various \\
Ballroom \cite{BallroomData}&$698 - e$& 12219 & 6h 04m & Ballroom dances\\
Quaero \cite{Quaeroweb} &$70$& 7104 & 2h 46m & Pop, rap, electro \\
Beatles \cite{BeatlesData} &$179$& 13937 & 8h 01m & Pop \\
RWC Pop \cite{Goto2002rwc}&$100$& 10835 & 6h 47m & Pop\\
\hline
Total & 1511 & 78171 & 43h 05m & \\
\end{tabular}}
\label{dataset_table}
\end{center}
\end{table}
}

\subsection{Comparative analysis}
\label{ssec:comparison}
We compare our system, called here ACNN\footnote{A for all features used, CNN for the learning method}, to the downbeat trackers of Peeters et al.~\cite{Peeters2011}, Davies et al.~\cite{Davies2006}, Papadopoulos et al.~\cite{Papadopoulos2011DB} and Krebs et al.~\cite{Krebs2015ismir}. It is worth noting that the train set of~\cite{Krebs2015ismir} overlaps with the Hainsworth dataset and that the algorithm of~\cite{Davies2006} was manually fed with the ground truth time signatures since it was needed to run it.  
Results are shown in Table~\ref{sota_table} and highlight an important increase by our system for each dataset and an overall increase of 16.8 percentage points (pp) compared to the second best system. The relative difference of the pop music datasets is lower at 11.1 pp because other algorithms are doing a fairly good estimation already. The harmonic change, spectral energy distribution and rhythmic patterns assumptions made by the compared systems are well followed in those datasets. However, in the other music datasets where the downbeats are more difficult to obtain, the overall increase in performance is much higher, at 18.9 pp. Our more sophisticated feature extraction and learning model is indeed doing a good job on some excerpts where downbeat cues are harder to obtain.\\

A particularly interesting case is the ballroom dataset. There is indeed a whooping 27.6 pp difference in performance compared to the second best system. Our system has a similar performance with this dataset and the pop music datasets while the compared algorithms have a much lower performance here. There is a lot of triple meter songs here, about 30\%. For comparison, there are only 2\% of triple meter in the RWC Pop dataset. However, it does not seem to be the reason on the difference of performance because our algorithm does not handle triple meter songs better than the compared systems, compared to duple meter songs. The difference may then be explained by the fact that the explicit rhythmic assumption of~\cite{Peeters2011} and~\cite{Krebs2015ismir} are not well verified. The performance of~\cite{Krebs2015ismir} increases drastically if it uses rhythmic patterns really close to the one used in this dataset~\cite{Krebs2013}.

Besides, the assumption of~\cite{Davies2006},~\cite{Papadopoulos2011DB} and~\cite{Peeters2011} that the harmonic content is different before and after a downbeat position may not be well verified here. To assess this, we used the chroma input only and replaced our harmonic network with a heuristic function as in~\cite{Davies2006} to obtain the downbeat likelihood. We compared the results across datasets and found a poorer performance of -8 pp on the ballroom dataset compared to the other datasets. On the other hand, our harmonic network alone have a performance on the ballroom dataset comparable to the one obtained with the pop music datasets, as can be seen in Table~\ref{sota_table}. The network seems able to go beyond simple harmonic change rules and detect important harmonic events such as the start or the end of a musical sentence. 
Short melodic or harmonic events can be important to find downbeats and they tend to be lost during the average process of the chroma vectors over a longer period of time that is done in several methods, but the harmonic network can take them into account. Our network also seems less prone to noisy events because the content of a 9 tatum window is taken into account to estimate a downbeat position.\\ 

Davies et al. system seems to work well on songs with correct beat estimation but it uses pre-estimated beat positions and an hypothesis of constant time signature that can propagate errors easily. Its performance increases significantly with the use of a more powerful beat tracker such as the one of Degara et al~\cite{Degara2012} with an overall F-measure of 56.7\%. At the opposite, \cite{Papadopoulos2011DB} is adapted to changes in the time signature but may be a bit too sensitive to these changes. \cite{Krebs2015ismir} includes a sophisticated temporal model but uses rhythmic features only. \cite{Peeters2011} performs a global estimation of the meter with an efficient visualization of the output and may be improved with a feature deep learning stage to be less dependent on certain downbeat assumptions.\\ 

The performance of our algorithm can be separated in three main categories, highlighted by horizontal lines in the Table~\ref{sota_table}. First is the RWC Class dataset with a rather low F-measure of about 50\%. In this case, the tatum estimation is uncertain and it is therefore difficult to estimate the downbeat likelihood or use the same temporal model as for the other styles. Besides, annotation is more difficult to perform, especially with soft onsets and romantic pieces, and the $\pm$70 ms tolerance window of the evaluation measure may be too short in many cases. A better tatum segmentation will significantly improve the results, as will be shown in section~\ref{sssec:segTest}. Finally, listening tests show that some classical music pieces are inherently more difficult to estimate without additional information even for an expert listener. Second, the Ballroom, Quaero, Beatles and RWC Pop datasets can be regrouped with a F-measure between 80\% and 90\%. These datasets contain a stable tempo easy to be tapped and are often not surprising in terms of their metrical structure or cues in order to infer the downbeat position and are well estimated by our system. Finally, a third set can be composed of the Klapuri subset, Hainsworth, RWC Jazz and RWC Genre datasets with a F-measure between 65\% and 70\%. Most of them are composed of a mixture of different genres and therefore have a performance in between the ones that are difficult or easy to estimate. The RWC Jazz dataset also belongs in this category because on one hand, the songs there contain for the most part clear rhythmic events and a relatively stable tempo. On the other hand, several instruments are sometimes playing more freely rhythmically, especially during solos. Finally many tempo errors occur because it is harder to define the correct metrical level. The F-measure is indeed increased by 10 pp if double or half tempo variations are allowed. It is the highest increase of all datasets.\\

The standard deviation of the F-measure across songs for this task is high but varies across databases. In our case, the mean of the standard deviation across songs, datasets and algorithms is 30\%.
The F-measure can indeed easily go from 100\% to 0\% if the third beat of the bar is taken for the downbeat for example, as it happens frequently. The standard deviation is lower for the RWC Class dataset because the time signature changes more regularly and the downbeat estimation is less consistent, limiting the 100\% and 0\% occurrences. At the opposite, the standard deviation is higher on the datasets composed of short excerpts, up to 40\%, because missing a downbeat has a higher effect on the performance. 


{\footnotesize
\begin{table}
\begin{center}
\caption{\footnotesize F-measure results for several published downbeat trackers.}
\begin{tabular}{ llllll}
  & Peeters & Davies &  Papadopoulos & Krebs & \textbf{ACNN}\\
  & et al. & et al. & et al. & et al. & \\
\hline\hline

RWC Class &29.9 & 21.6 & 32.7 & 33.5 & \textbf{51.0}\\
\hline

Hainsworth & 42.3 & 47.5 & 44.2 & 51.7* & \textbf{65.0}\\

RWC Genre & 43.2 & 50.4 & 49.3 & 47.9 & \textbf{66.1}\\

Klapuri &47.3 & 41.8 & 41.0 & \textbf{50.0} & \textbf{67.4}\\

RWC Jazz & 39.6 & 47.2 & 42.1 & 51.5 & \textbf{70.9}\\
\hline

Ballroom & 45.5 & 50.3 & 50.0 & 52.5 & \textbf{80.1}\\

Quaero & 57.2 & 69.1 & 69.3 & 71.3 & \textbf{81.2}\\

Beatles & 53.3 & 66.1 & 65.3 & 72.1 & \textbf{83.8}\\

RWC Pop & 69.8 & 71.0 & 75.8 & 72.1 & \textbf{87.6}\\
\hline

\textbf{Mean}& 47.6 & 51.7 & 52.2 & 55.8 & \textbf{72.6}\\

\end{tabular}
\label{sota_table}
\end{center}
\end{table}
}


\subsection{Detailed analysis of the proposed system}
An analysis of each step of the proposed system is given below.

\subsubsection{Segmentation} How much do the limitations of our tatum segmentation affect the performance?
\label{sssec:segTest}
The tatum segmentation step, with the advantage of segmenting the data into a reduced set of rhythmically meaningful events, has a downbeat recall rate of 92.9\% considering a $\pm$70 ms tolerance window and therefore occasionally misses an annotated downbeat. How much can it affect the overall performance? To assess this, we kept the system the same and only replaced the closest estimated tatum position to a ground truth downbeat by its annotated position in the segmentation step to obtain a perfect recall rate. Mean results across datasets are shown in the row (a).1 of the figure~\ref{fig:resAllFig} and we see an improvement of 3.9 pp compared to the reference model. This is statistically significant. Results in the RWC Pop, Quaero and Ballroom datasets of systems with or without a perfect downbeat recall are fairly close with a relative difference of about 1.4 pp. It highlights that our tatum segmentation step is very reliable to detect downbeats in music with a clear rhythmic structure. Improvement of 4.1 pp in the Beatles dataset may be contradictory, but a subjective error analysis shows that most of the difference in performance there is due to a questionable annotation. However, for other genres, a bigger difference of 5.4 pp appears, highlighting some of the limitations of the timing of estimated tatums near downbeat positions.\\ 










The segmentation step can have other issues than imprecise timing around downbeat positions. For example, two consecutive bars may contain a different number of estimated tatums. It also happens that tatums don't have the right periodicity or do not match downbeats at all. To assess this effect without changing the rest of the system, we want to use the best tatum segmentation for a given metrical level. Since we have access to annotated beats but not to annotated tatums, we replace the tatums with an interpolation by a factor 2 of the annotated beats in the segmentation step. This is the most common factor between beats and tatums in our datasets. 
Results are shown in the row (a).2 of the figure~\ref{fig:resAllFig} and we see an improvement of 11.4 pp. This is statistically significant. The overall increase in performance is much higher here because all tatum positions are modified to match the downbeat sequence instead of only those near downbeats in the former experiment. The segmentation is therefore a lot cleaner and octave error are a lot less common. Compared to the perfect downbeat recall case, the improvement is particularly striking on the RWC Jazz dataset (9.4 pp) and the RWC Classical dataset (20.4 pp). Many octave errors were present in the first case and are now solved. In the second case, many spurious events and a lack of tatum consistency inside a bar were greatly disturbing the system. The hard part in the Classical music dataset is mostly to have a proper sub-downbeat segmentation. It is to note that the overall performance when beats are known is a bit overfitted to the annotations and may not be a precise estimate of the performance of our system with a better segmentation step. Indeed, besides some human errors in annotation, this task is sometimes subjective in terms of downbeat timing and metrical level. However, it appears that designing a more precise, clean and consistent segmentation step may have a significant effect on the overall performance.\\ 

\subsubsection{Feature adapted neural networks} Are all feature adapted neural networks useful?

Using several complementary features and adapted learning strategies is the core of our method. However, there is a limit to the added value of a new feature compared to its redundancy with the others and not all features may be worth adding in our model, especially when using the average as a feature combination. 

\begin{figure}

  \centering
\includegraphics[width=8.7cm]{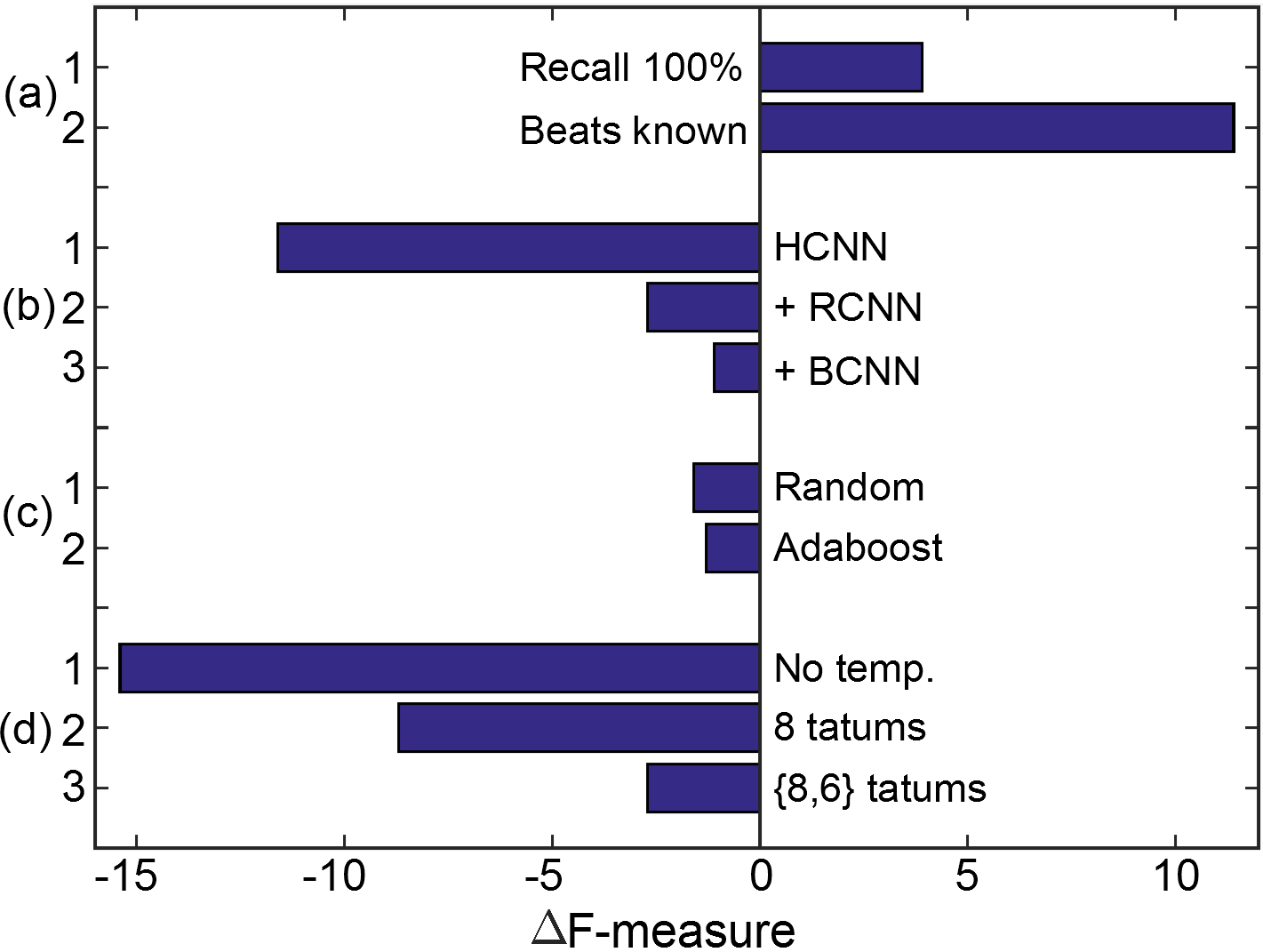}
\caption{\footnotesize Relative F-measure for various configurations of our system compared to the standard model (ACNN). Mean across all datasets. (a): Segmentation variation. (b): Feature variation. (c): Network combination variation. (d): Temporal model variation.}
\label{fig:resAllFig}
\end{figure} 

To assess the effect of each feature adapted neural network efficiently, we show the performance of the best performing network, followed by the performance of the best combination of two networks, and then the performance of the best combination of three networks compared to our four networks system in figure~\ref{fig:resAllFig}(b). The harmonic neural network, figure~\ref{fig:resAllFig}(b).1, is quite powerful by itself with an overall F-measure of 61.0\%, that is already significantly better than the compared algorithms in subsection~\ref{ssec:comparison}. On a dataset strongly depending on harmony such as the RWC Classical dataset, the performance of the harmonic network only is close to the one obtained with the whole model with a 4.7 pp difference. However, using the four networks is more robust on a variety of datasets and leads to a 11.6 pp increase overall. The harmonic and rhythmic networks is the best two-network configuration with an overall F-measure of 69.9\%. Harmonic and rhythmic features are indeed quite complementary in a wide range of music. However, a statistically significant improvement is achieved by adding the bass content network with an overall F-measure of 71.5\% and also by using all the features with an overall F-measure of 72.6\%. 

While the system complexity increases with each new network, it remains relatively low. The mean processing time for 60 seconds of musical audio is respectively of 1.0, 1.5, 2.0 and 12.5 seconds as we add features and their corresponding network as in figure~\ref{fig:resAllFig}(b) on a PC with an Intel Xeon e5-1620 CPU with 3.6 GHz. The melodic network increases the processing time by a larger margin because of its bigger input. The training time of all four networks is about one day on a Geforce GTX TITAN Black GPU.\\

\subsubsection{Feature learning} Is deep learning useful?

Considering that the combination of the harmonic and rhythmic networks provides a very good performance already and for complexity and clarity constraints, we will restrain the experiments in this part to these two networks. We will then consider several ways of getting a downbeat likelihood from our chroma and onset detection function inputs individually and in combination. We will compare a shallow learning method, a deep learning method and finally the hereby presented feature adapted and locally dependent deep learning.

{\footnotesize
\begin{table}
\begin{center}
\caption{\footnotesize Mean F-measure results for various feature processing steps.}
\begin{tabular}{llll}

Inputs used & SVM & DBN & CNN\\
\hline \hline
ODF & 49.0 & 53.9 & \textbf{57.0}\\

Chroma & 43.6 & 52.2 & \textbf{61.1}\\

ODF + Chroma & 56.6 & 64.5 & \textbf{69.9}\\

\end{tabular}
\label{feature_proc_table}
\end{center}
\end{table}
}

We first use linear Support Vector Machine (SVM) with a penalty parameter $C$ to the error term found on the validation set as our shallow learning method. As SVM predicts only class labels, the probability of each class will be estimated following the second method of~\cite{Wu2004prob} that is based on Platt scaling. As for the harmonic network, we used the same number of downbeat and no downbeat examples in our training data to have a balanced training set. There is also one classifier per input type and the fusion of several classifier is done by averaging. Results, presented in Table~\ref{feature_proc_table} show that the rhythmic SVM is significantly better than the harmonic SVM and that their combination is competitive with the algorithms presented in subsection~\ref{ssec:comparison}\footnote{A better SVM model, adapted feature, or probability estimate may be found but this is not the focus of this work.}.

We will now take advantage of the relatively large number of training examples and the high level aspect of downbeats to apply a relatively simple deep learning method to our problem. We use Deep Belief Network (DBN) as in~\cite{Durand2015} as our deep learning method. The used networks consist of three fully connected layers of 40, 40 and 50 units respectively, a sigmoid activation function and a softmax regularization output. Results in Table~\ref{feature_proc_table} highlight a significant improvement for both features and their combination. However, we see in Table~\ref{feature_proc_table} that using locally dependent feature adapted networks is more suited to our problem and also significantly improves the results. Finally using both inputs leads to significantly better results in all cases.\\

\subsubsection{Downbeat likelihood combination} Are more sophisticated combinations useful?

We are currently using the average of the downbeat likelihood computed with each of the four networks to obtain one single estimation. Since this method may seem too simple, we provide a comparison with two other feature combination techniques.
We first compare our fusion method with the Adaboost algorithm~\cite{Freund1999}. A linear combination of the classifiers will be learned by emphasizing the ones that correctly classify an instance mis-classified by the other classifiers. This approach can work well in practice but it is better suited to a problem involving many weak classifiers. We use a learning rate of 0.10 and an ensemble of 100 trees.

Random forests are also tested. A multitude of decision trees are constructed by this ensemble learning method, to predict the class that is chosen by most individual trees~\cite{Breiman2001}. The probability output is computed as the number of observations of the selected class in a tree leaf over the number of trees in the ensemble. As deep learning algorithms, this method often requires a good number of training examples to work well. We use 30 trees and a leaf size of 50. For those two methods, as we did with the harmonic network, we randomly remove some training inputs in order to have a balanced training set.



Results are shown in figure~\ref{fig:resAllFig}(c). We can see a decrease of 1.3 pp with boosting and of 1.6 pp with random forest overall. This is statistically significant. This result may seem surprising at first, but may be explained by the fact that those two algorithms minimize classification error on the training set for each instance individually, while we evaluate the F-measure on full songs after the temporal model. Adaptation to the temporal modeling phase is therefore key and neither boosting nor random forest focus on this part. We found that the boosting method has a similar performance for the three Pop music datasets but is less robust to other sets. The average rule is indeed often more resilient and will not overfit a particular and more represented set. Besides, since we only have 4 features that are relatively strong, comparable in performance and complementary, an average of the result can give a good result already. 

\subsubsection{Temporal model} Is the temporal model useful?

The temporal model is an important part of our system as it allows an increase in performance of 15.4 pp as can be seen in the figure~\ref{fig:resAllFig}(d). In the configuration without the temporal model, figure~\ref{fig:resAllFig}(d).1, a downbeat position is included in the final downbeat sequence if its likelihood is above a fixed oracle threshold. This oracle threshold $t=0.88$ was manually selected to give the best F-measure result. It corresponds roughly to the ratio of downbeats and no downbeats in the datasets. 

The important gap in performance can be explained by the fact the downbeat likelihood function is quite noisy as can be seen at the bottom of figure~\ref{fig:hcnn}. Besides, 9 or 17 tatum long inputs are sometimes too short to give a reliable information about the downbeat position by themselves. It is therefore the heavily structured nature of music that enables the system to obtain a more sensible downbeat sequence.

However, when a simple temporal model of only 8 states (for an 8 tatum bar) is considered, the performance increases considerably as seen in figure~\ref{fig:resAllFig}(d).2. And when 14 states (for 8 and 6 tatum bars) are considered, the F-measure performance is not very far from the one obtained with our full 80 states model as seen in figure~\ref{fig:resAllFig}(d).3. It highlights that our segmentation step that is relatively consistent, but also a lack of diversity in the used datasets time signatures. The performance gap is low because more than 99\% of the songs are mainly in 2, 3 or 4 beats per bar. 
However, at the moderate cost of only 18 ms per one minute song that is two orders of magnitude lower than the whole processing time, the 80 states temporal model gives a significantly better performance of +2.7 pp. 
Downbeat sequences of three and four beats per bar tracks are fairly well estimated with a mean F-measure of 79.5\% and 77.4\% respectively. The performance of 2 beats per bar tracks is significantly lower, at 55.0\% since they are mostly taken for 4 beats per bar tracks. This is a common ambiguity for algorithms and human listeners that goes beyond the scope of this work.



\section{Conclusion and future work}
In the present work, we have presented a system that robustly detects downbeat occurrences from various audio music files. Our work is based on the fact that a downbeat is a high level concept depending on multiple musical attributes. We therefore have designed complementary features based on harmony, melody, bass content and rhythm and adapted convolutional neural networks to each feature characteristics. Rhythm in music being highly structured, an efficient and flexible temporal model was also proposed to largely improve the instantaneous downbeat estimation. A comparative evaluation on a large database of 1511 audio files from various music styles shows that our system significantly outperforms four other published algorithms, while keeping a low computational cost. Each step of our system was analyzed to highlight its strengths and shortcomings. In particular, the combination of harmonic and rhythmic deep networks proved to be very good by itself.

While the proposed algorithm obtained the best results overall, it is to note that the recent MIREX campaign\footnote{\url{http://www.music-ir.org/mirex/wiki/2015:Audio_Downbeat_Estimation_Results}} highlighted some limitations in our method. The submitted system does not match exactly the one presented here but is based on the same framework. 
First, the temporal model doesn't deal well with 2 beats per bar songs as in Cretan music for example as it can easily be confused with 4 beats per bar songs. This issue can be fixed by adapting the temporal model to the music convention of a particular style. Second, while it can adapt to music of different traditions such as Turkish Usuls fairly well, music coming from a much more different genre such as Indian Carnatic or with particular rhythm conventions such as Hardcore, Jungle and Drum and Bass requires a training set containing some adapted examples for the system to work better. These remarks are rather expected as a human listener hearing these music styles for the first time will also tend to be lost before training~\footnote{Several audio excerpts are available at \url{http://www.music-ir.org/mirex/wiki/2015:Audio_Downbeat_Estimation} for the interested listener.}. It shows that bars are not intuitively understood for all music traditions and that designing a more adapted or exhaustive training set is important. However, the number of music tracks for which our system provides a good estimation is still important.

In the future, besides a more refined training set, a network combination procedure adapted to the temporal model and a more robust segmentation step seem promising to improve the current system.


%







\bibliographystyle{IEEEtran}
\bibliography{sample}
\end{document}